\documentclass[a4paper,11pt]{article}
\usepackage{pos}
\usepackage{hyperref}
\usepackage{graphicx}
\usepackage{mathtools}
\usepackage{float}
\usepackage[english]{babel}
\usepackage[autolanguage]{numprint}

\newcommand{\nucl}[3]{\mbox{$\prescript{#1}{#2}{\rm #3}$}}
\newcommand{\pw}[3]{\mbox{$\prescript{#1}{}{\rm #2}_{#3}$}}

\title{Constraints on the $\Lambda$-neutron interaction from 
charge-symmetry breaking of $A=4$ hypernuclei}
\ShortTitle{Constraints on the $\Lambda n$ interaction from 
CSB of $A=4$ hypernuclei}

\author*[a,b]{Andreas Nogga}
\emailAdd{a.nogga@fz-juelich.de}
\author[a]{Johann Haidenbauer}
\author[c,a,b]{Ulf-G. Mei{\ss}ner}

\affiliation[a]{IAS-4, IKP-3, and JHCP, Forschungszentrum J{\"u}lich, D-52425 J{\"u}lich, Germany}
\affiliation[b]{CASA, Forschungszentrum J{\"u}lich, D-52425 J{\"u}lich, Germany}
\affiliation[c]{HISKP and BCTP, Universit\"at Bonn, D-53115 Bonn, Germany }

\abstract{We include the leading charge symmetry breaking 
contributions into the hyperon-nucleon interactions 
derived within chiral effective field theory 
up to next-to-leading order. Two low energy constants are determined using the experimentally known differences of $\Lambda$ separation 
energies of \nucl{4}{\Lambda}{He} and \nucl{4}{\Lambda}{H}. 
This allows one to predict the $\Lambda$-neutron scattering lengths 
for the first time based on data. Various sources of uncertainty 
are discussed. }

\FullConference{%
  The 10th International Workshop on Chiral Dynamics - CD2021\\
  15-19 November 2021\\
  Online
  }

\begin{document}
\maketitle

\section{Introduction}
\label{sec:intro}

To understand the interior of neutron stars a thorough 
knowledge on the equation of state (EOS) of nuclear matter is 
the prerequisite. This includes to also pin down the possible 
contribution of hyperons to nuclear, especially neutron, matter. 
Simply including hyperon degrees of freedom  
leads to a much too soft EOS that does not support the existence 
of neutron stars with mass about or larger as two solar masses
\cite{Gandolfi:2015jma,Tolos:2020aln,Chatterjee:2015pua,Weissenborn:2011ut}. This is generally referred to as hyperon puzzle.
For a microscopic understanding of nuclear matter, it is therefore 
of utmost importance to study the properties of 
hyperon-nucleon (YN) interactions. Such YN interactions should then not only describe all available YN scattering data and 
hypernuclei but also lead to a contribution of hyperons to 
nuclear matter that is consistent with the observed neutron star
masses \cite{Demorest:2010bx,Antoniadis:2013pzd,Fonseca:2016tux,NANOGrav:2017wvv,NANOGrav:2019jur} and radii \cite{Riley:2019yda,Miller:2019cac,Riley:2021pdl,Miller:2021qha}. 

Obviously, the most relevant interaction to be studied in this context is the one of $\Lambda$ and neutron ($\Lambda n$). Unfortunately, there is no scattering data for elastic $\Lambda n$
scattering so that the interaction is usually determined by $\Lambda $ proton ($\Lambda p$) data assuming isospin symmetry. 
Very early on, it was observed that isospin symmetry is significantly violated 
for these systems as can be seen from the charge symmetry breaking (CSB) 
$\Lambda$ separation energy differences of $A=4$ hypernuclei. Depending on the state considered,
it can be as larger as 250~keV which is at least four times larger than the strong 
interaction contribution to the CSB of $A=3$ ordinary nuclei \cite{Nogga:2002qp}.  
One important contribution to this CSB has been related to $\Sigma^0$-$\Lambda$ mixing. 
This effectively results in a one-pion exchange contribution to the $\Lambda$N 
interaction which is CSB \cite{Dalitz:1964es}. The contribution is part of, e.g., the 
Nijmegen SC97 interactions \cite{Stoks:1999bz}
but cannot explain the  CSB of $A=4$ hypernuclei within this model \cite{Nogga:2001ef}. 
Gal and Gazda \cite{Gal:2015bfa,Gazda:2015qyt,Gazda:2016qva} recently studied the same contribution  based on the leading order (LO) chiral
YN interaction \cite{Polinder:2006zh}. While they found that for specific choices 
of the cutoff the CSB is correctly reproduced, they also observed 
a considerable dependence on the cutoff.

Besides the phenomenological importance, the CSB is also conceptually interesting
since it is linked to the $\Lambda$-$\Sigma$ conversion of YN interactions. 
One important contribution to the CSB of $A=4$ hypernuclei is related to the 
mass difference of $\Sigma$ particles. The strong $\Lambda$-$\Sigma$ conversion 
process of YN interactions leads to a small, unfortunately non-observable, $\Sigma$ 
component of hypernuclear wave functions. Model calculations show that this mass differences 
leads to a visible contribution to the CSB of the kinetic energy which is essentially 
proportional to the $\Sigma$ probability for the hypernucleus \cite{Nogga:2001ef}. 

From the perspective of chiral effective field theory, 
there are two momentum 
independent, CSB contact interaction at the same order as the CSB one-pion exchange.
So far, these contact interactions have always been neglected. 
In Ref.~\cite{Haidenbauer:2021wld}, we introduce these contact interactions for the first time
and determined the values of the related low energy constants (LECs) 
using the experimentally known values 
for the splitting of the separation energies of $^4_\Lambda$H and $^4_\Lambda$He  
in the $0^+$ and $1^+$ states. This completely determines the leading CSB interaction 
and allows one to predict  $\Lambda n$ and $\Lambda p$  singlet and triplet scattering lengths based on data. In this contribution, we summarize these results. We start with a short discussion
on chiral YN interactions in Sec.~\ref{sec:ynint}  focusing on our approach 
to estimate higher order contributions, especially three-baryon forces (3BFs). 
In Sec.~\ref{sec:csbyn}, we discuss the most important CSB contributions to YN interactions
based on chiral effective field theory (EFT). The 
determination of the two CSB LECs and predictions for the 
$\Lambda$N scattering lengths are shown in Sec.~\ref{sec:res}. Finally, we conclude and give 
an outlook in Sec.~\ref{sec:concl}. 

\section{Chiral YN interactions and estimates for 3BF contributions}
\label{sec:ynint}

Our study is based on chiral YN interactions at next-to-leading order (NLO) \cite{Haidenbauer:2013oca, Haidenbauer:2019boi}. 
Up to this order, there are contributions of one- and two pseudoscalar (Goldstone) boson 
exchanges and more than 20 different short range contact interaction. 
Coupling constants of Goldstone bosons to baryons can be related by using SU(3) flavor symmetry. On the other hand, SU(3) symmetry 
is broken by using physical masses 
for the octet mesons. Since there are only 35 data available (new data has been obtained only recently \cite{J-PARCE40:2021bgw,J-PARCE40:2021qxa}), a unique determination of all low energy constants is not possible. In fact, it turned 
out that it is even possible to define two realizations of the chiral 
interaction at NLO that are almost phase shift equivalent. We refer to these 
version as NLO13 \cite{Haidenbauer:2013oca} and NLO19 \cite{Haidenbauer:2019boi}. 
These chiral YN interactions need to be regularized. For both NLO13 and NLO19, 
a cutoff between 500~MeV and 650~MeV is used for this regularization. 
Within this range of cutoffs, we obtained the best description of the 
available YN data.

\begin{table}
\begin{center}
    
\begin{tabular}{l|r|rr|rr}
      interaction &  $E_\Lambda(^3_\Lambda{\rm H})$ &  \multicolumn{2}{c|}{$E_\Lambda(^4_\Lambda{\rm He})$ } &  \multicolumn{2}{c}{$E_\Lambda(^4_\Lambda{\rm H})$ }\\
                     &     &   $J^\pi=0^+$ & $J^\pi=1^+$ & $J^\pi=0^+$ & $J^\pi=1^+$ \\                                      
\hline
NLO13(500)           & 0.13 &  1.71 &  0.80 &  1.66 & 0.78  \\
NLO13(550)           & 0.09 &  1.51 &  0.59 &  1.45 & 0.57   \\
NLO13(600)           & 0.09 &  1.48 &  0.59 &  1.43 & 0.56   \\
NLO13(650)           & 0.08 &  1.50 &  0.62 &  1.45 & 0.60   \\
\hline
NLO19(500)           & 0.10 &  1.65 &  1.23 &  1.63 & 1.23   \\
NLO19(550)           & 0.09 &  1.55 &  1.25 &  1.53 & 1.24   \\
NLO19(600)           & 0.10 &  1.47 &  1.06 &  1.44 & 1.05   \\
NLO19(650)           & 0.09 &  1.54 &  0.92 &  1.50 & 0.91   \\
\hline 
Expt.                & $0.13(5)$ \cite{Juric:1973zq} 
      & 2.39(3)\cite{J-PARCE13:2015uwb} & 0.98(3)\cite{J-PARCE13:2015uwb} 
      & 2.16(8)\cite{A1:2016nfu} & 1.07(8) \cite{A1:2016nfu} \\
\end{tabular}
\end{center}

\caption{\label{tab:h3lamhe4lamh4lam} 
$^3_\Lambda$H, $^4_\Lambda$He and $^4_\Lambda$H
separation energies for NLO13 
and NLO19 for various cutoffs. No explicit CSB is included in the $YN$ 
 potentials. Energies are in MeV. 
}
\end{table}

As has been discussed in Ref.~\cite{Haidenbauer:2019boi}, both version 
differ mainly by their strength of the $\Lambda$-$\Sigma$ transition 
potential thereby keeping the transition cross sections very similar to each 
other. Although, both interactions give almost identical results for the 
YN system, the predictions for systems with $A>2$ are different. Such 
differences can be traced back to higher order contributions, i.e. three-baryon
force (3BF) contributions, which first appear 
at next-to-next-to-leading order (N$^2$LO)
\cite{Petschauer:2015elq}. 

In Table~\ref{tab:h3lamhe4lamh4lam}, this 
is shown for $A=3$ and $A=4$ hypernuclei. For calculating these systems, we also have to 
employ NN interactions. Since we will be discussing $\Lambda$ separation 
energies, i.e. the differences of the binding energy of the core and 
the hypernucleus, the results will only mildly dependent on the 
NN interaction used \cite{Haidenbauer:2019boi,Nogga:2001ef}. In this 
work, we therefore only employed one NN interaction: the chiral SMS 
regulated interaction at order N$^4$LO+ with a cutoff of 450~MeV \cite{Reinert:2017usi}.  

It is reassuring that the variation 
of $\Lambda$ separation energies is only about 40~keV for \nucl{3}{\Lambda}{H}. 
This is of the order of the experimental uncertainty and justifies to 
adjust the relative strength of the \pw{1}{S}{0}  and \pw{3}{S}{1} 
YN interactions such that \nucl{3}{\Lambda}{H} is described in agreement with 
experiment. In absence of YN data determining the spin dependence, this 
essentially fixes the $\Lambda$N scattering lengths. Note that this 
approach has only been done for the 600~MeV cutoff. For the other cutoffs, 
the interaction has been fitted to the scattering length obtain for 600~MeV. 
Therefore, the small variation of the energies are still a measure 
for N$^2$LO, especially 3BF contributions to \nucl{3}{\Lambda}{H}. Due 
to the small separation energy of this system, such contributions 
seem to be suppressed. 

This is not case for $A=4$ hypernuclei anymore.  For the $0^+$ state, 
the variation of the $\Lambda$ separation energy is about 240~keV. For the 
$1^+$ state, it is even as large as 650~keV. Clearly, eventually, 3BFs need to 
be added to improve the predictions for these hypernuclei. However, for a first 
study of CSB in $A=4$ hypernuclei, the description is still 
sufficiently realistic to allow for a sensible estimate. It will be important 
later on to perform the calculations using both versions of the NLO 
interactions and all cutoffs to ensure that higher order contributions 
do not affect the results significantly. 

The results shown in Table~\ref{tab:h3lamhe4lamh4lam} do not include CSB 
of the YN interaction. But they already include CSB due to the mass differences 
of $\Sigma^+$, $\Sigma^0$, $\Sigma^-$ \cite{ParticleDataGroup:2020ssz}. 
Also the contribution of the proton-proton Coulomb interaction changes compared to the one of the core nucleus. The inclusion of these two effects explains 
the small CSB of the separation energies 
shown in the table.

\section{Leading contributions to CSB YN interactions}
\label{sec:csbyn}

\begin{figure}[t]
\begin{center}
\includegraphics[height=2.5cm]{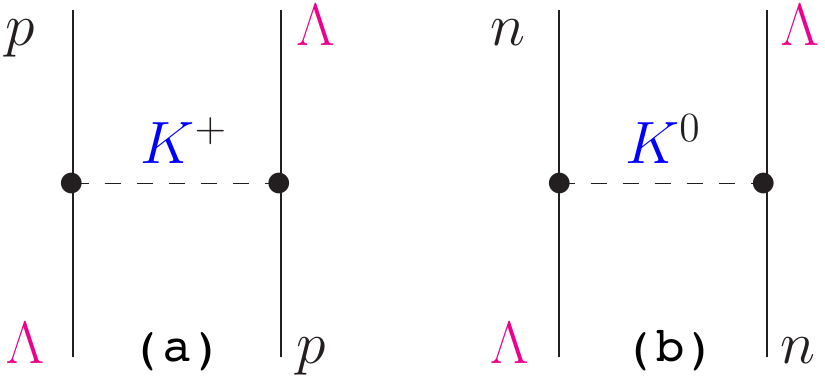} \hspace{1cm}
\includegraphics[height=2.5cm]{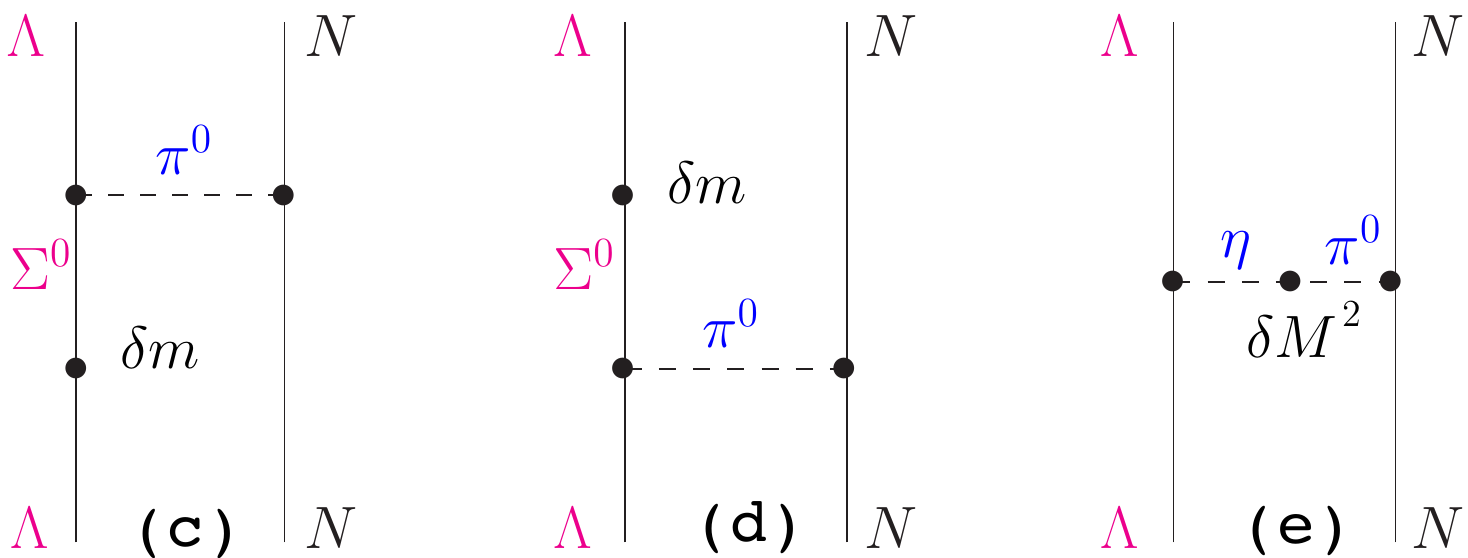}
\caption{CSB contributions involving kaon and pion exchange.}
\label{fig:csbdiagrams}
\end{center}
\end{figure}

The implementation of CSB in chiral EFT has already been discussed for the 
nucleon-nucleon (NN) system in Refs.~\cite{Walzl:2000cx,Friar:2003yv,Epelbaum:2004fk}.
Thereby, the relevant expansion parameter is 
$\epsilon M^2_\pi / \Lambda^2 \sim 10^{-2}$, where 
$\epsilon \equiv \frac{m_d-m_u}{m_d+m_u} \sim 0.3$ is a ratio of quark masses and 
$\Lambda\sim M_\rho$. 

The formally leading contributions at order $n=1$ (L\O) are due to the Coulomb interactions  and 
due to mass differences between the exchanged pseudo-Goldstone bosons, i.e. 
$M_{\pi^\pm}-M_{\pi^0}$ and $M_{K^\pm}-M_{K^0}$. For the $\Lambda$N system, 
only the contributions due to $M_{K^\pm}-M_{K^0}$ contribute (see Fig.~\ref{fig:csbdiagrams} (a) and (b)). Because the kaon masses are rather large compared to their mass
difference, it turns out that this effect is actually very small. It is included in our 
calculations, but is quantitatively not important. The more 
important contributions are therefore formally subleading of order $n=2$ (NL{\O}). 
This includes the isospin violation in the pion-baryon coupling constant due to 
$\Sigma^0-\Lambda$ mixing as well as from $\pi^0-\eta$ mixing leading to 
a CSB one-pion exchange contribution to the $\Lambda$N interaction as depicted 
in Fig.~\ref{fig:csbdiagrams} (c) to (e). The former 
mechanism was already introduced in Ref.~\cite{Dalitz:1964es} and is the basis 
of most studies of CSB in the YN system. 
Using the PDG mass values~\cite{ParticleDataGroup:2020ssz}, one obtains
for the CSB effective $\Lambda$N coupling constant 
\begin{equation}
\label{eq:Hippel}
f_{\Lambda\Lambda\pi} = \left[
-2 \frac{\langle \Sigma^0|\delta m|\Lambda\rangle}{m_{\Sigma^0}-m_{\Lambda}}+ 
\frac{\langle \pi^0|\delta M^2|\eta\rangle}{M^2_{\eta}-M^2_{\pi^0}}
\right]\, f_{\Lambda\Sigma\pi} \approx (-0.0297 - 0.0106)\, f_{\Lambda\Sigma\pi} .
\end{equation}

\begin{figure}[t]
\begin{center}
\includegraphics[height=2.5cm]{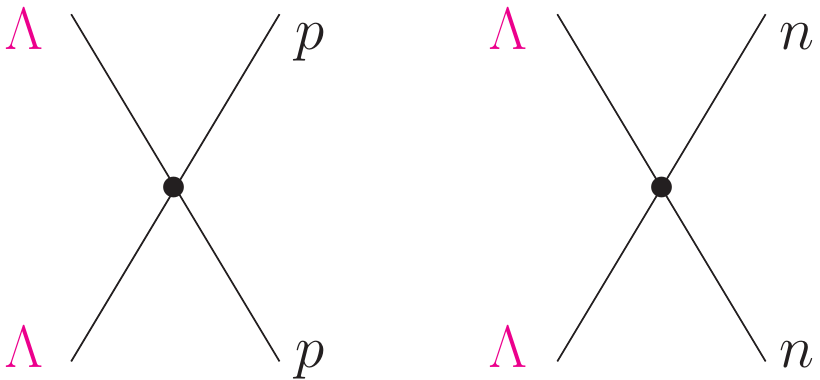} 
\caption{CSB contact interactions contributing to $\Lambda$N interactions.}
\label{fig:csbct}
\end{center}
\end{figure}

New in our work is that we take into account the additional contributions from short range forces (arising from, e.g., $\rho^0-\omega$ mixing) which, in chiral EFT, are represented by contact terms involving CSB LECs (see Fig.~\ref{fig:csbct}).
There are two such contact forces, one for singlet and one for triplet $\Lambda$N, which 
are also required to properly renormalize the one-pion exchange contribution. In the following, we will determine these 
LECs using the CSB splitting of the separation energies of \nucl{4}{\Lambda}{H} and \nucl{4}{\Lambda}{He} for the 
$0^+$ and $1^+$ state. 

\section{Results}
\label{sec:res}

In order to predict the CSB for the $\Lambda$N scattering lengths, we first 
need to determine the CSB LECs for the singlet ($C^{CSB}_{s}$) and 
triplet ($C^{CSB}_{t}$) states. To this aim, we use the difference of the $\Lambda$ separation energies 
\begin{eqnarray}
\Delta E(0^+) & = & E^{0^+}_{\Lambda}(\nucl{4}{\Lambda}{He})-E^{0^+}_{\Lambda}(\nucl{4}{\Lambda}{H}), \nonumber \\
\Delta E(1^+) & = &  E^{1^+}_{\Lambda}(\nucl{4}{\Lambda}{He})-E^{1^+}_{\Lambda}(\nucl{4}{\Lambda}{H}) \ . 
\end{eqnarray}

Here, we only discuss results based on the central values of the 
present experimental situation, i.e. on the recent measurements of $E^{0^+}_{\Lambda}(\nucl{4}{\Lambda}{H})$ in Mainz \cite{A1:2016nfu}
and the one of $E^{0^+}_{\Lambda}(\nucl{4}{\Lambda}{He})-E^{1^+}_{\Lambda}(\nucl{4}{\Lambda}{He})$  at J-PARC \cite{J-PARCE13:2015uwb},
the old emulsion result of $E^{0^+}_{\Lambda}(\nucl{4}{\Lambda}{He})$  \cite{Juric:1973zq} and 
$E^{0^+}_{\Lambda}(\nucl{4}{\Lambda}{H})-E^{1^+}_{\Lambda}(\nucl{4}{\Lambda}{H})$ from \cite{Bedjidian:1976zh}. We refer to this 
scenerio as CSB1 leading to $\Delta E(0^+) = 233\pm 92$~keV and $\Delta E(1^+) = -83\pm 94$~keV. 
Note that we discuss in Ref.~\cite{Haidenbauer:2021wld}  also older experimental values for 
these splittings. 

The fitting is based on Faddeev-Yakubovsky calculations in momentum space using 
the isospin conserving YN interactions (see Table~\ref{tab:h3lamhe4lamh4lam}). 
We restricted all orbital angular momenta to $l\le 4$ and the YN and NN pair 
angular momentum also to $j \le 4$. Additionally, also the sum of 
the three orbital angular momenta required for our representation in Jacobi 
coordinates is restricted to 10. Thereby, only total isospin $T=1/2$ 
is taken into account. This insures that the numerical uncertainty is better than 
10~keV for the energies entering the Yakubovsky equations and 20~keV 
for expectation values of the energy. Interestingly, most of this uncertainty is due to 
the missing total isospin components with $T=3/2$ and $T=5/2$.

\begin{table}
\nprounddigits{3}
\begin{center}
\begin{tabular} {l|rr|rr}
    &\multicolumn{2}{|c|}{NLO13}  & \multicolumn{2}{|c}{NLO19} \\
\hline
$\Lambda$ &  $C_s^{CSB}$ &  $C_t^{CSB}$
&  $C_s^{CSB}$ &  $C_t^{CSB}$ \\
\hline
 500      &   \numprint{4.691413e-3}  &  \numprint{-9.294219e-4}   &  \numprint{ 5.589635e-3} &  \numprint{-9.505238e-4}  \\
 550      &   \numprint{6.724052e-3} &   \numprint{-8.625089e-4}   &   \numprint{6.863206e-3} &  \numprint{-1.260004e-3} \\
 600      &   \numprint{9.959831e-3}  & \numprint{-9.869969e-4}    &   \numprint{9.217360e-3}  & \numprint{-1.304647e-3} \\
 650      &   \numprint{1.500219e-2}  &   \numprint{-1.142164e-3}  &   \numprint{1.240454e-2}  & \numprint{-1.394845e-3} \\
\end{tabular}
\end{center}
\caption{\label{tab:LECs}CSB contact terms used for the singlet (s) and triplet (t) $\Lambda$N interactions. 
         The values of the LECs are in $10^4$~GeV$^{-2}$.}
\end{table} 

Based on the wave functions for \nucl{4}{\Lambda}{He}, we evaluated the individual 
contributions of the baryon mass differences in the kinetic energy, CSB one-pion exchange
and the singlet and triplet CSB contact interactions. This allows one to determine the 
corresponding LECs. The results are shown in Table~\ref{tab:LECs} and are 
well in line with the power counting expectation of 
$$ C_{s/t}^{CSB} \propto  \frac{m_d-m_u}{m_d+m_u} \left(\frac{M_{\pi}}{\Lambda}\right)^2 C_{s/t} 
\approx 6 \cdot 10^{-3} \ 10^4 \mbox{ GeV}^{-2} $$ 
where $C_{s/t}$ are the isospin conserving LO counter terms of the chiral $\Lambda$N interaction. 

\begin{table}
\begin{center}
\begin{tabular}{l|rrr|rr}
      interaction &  $\langle T \rangle_{{\rm CSB}}$ 
                  & $\langle V_{YN} \rangle_{{\rm CSB}}$
                  &  $V_{NN}^{\rm CSB}$ 
                  & $\Delta E_\Lambda^{pert}$ 
                  & $\Delta E_\Lambda$ \\
\hline
NLO13(500)       &  44 &  200 &   16 &  261 &   265 \\
NLO13(550)       &  46 &  191 &   20 &  257 &   261 \\
NLO13(600)       &  44 &  187 &   20 &  252 &   256 \\
NLO13(650)       &  38 &  189 &   18 &  245 &   249 \\
\hline
NLO19(500)       &  14 &  224 &    5 &  243 &   249 \\
NLO19(550)       &  14 &  226 &    7 &  247 &   252 \\
NLO19(600)       &  22 &  204 &   12 &  238 &   243 \\
NLO19(650)       &  26 &  207 &   12 &  245 &   250 \\
\end{tabular}
\end{center}
\caption{\label{tab:csbpert0} Perturbative estimate of the CSB 
of $^4_\Lambda$He  and $^4_\Lambda$H separation energies $\Delta E_\Lambda^{pert}$  for 
the $0^+$ state based on $^4_\Lambda$He wave functions compared to the full result $\Delta E_\Lambda$.
The individual contributions due to the kinetic energy $\langle T \rangle_{{\rm CSB}}$, the $YN$ interaction $\langle V_{YN} \rangle_{{\rm CSB}}$ and the contribution of the nuclear core   $V_{NN}^{\rm CSB} = \langle V_{NN} \rangle_{{\rm CSB}}- E(^3{\rm He}) + E(^3{\rm H})$ are also shown. All energies are in keV. }
\end{table}

The fits are then used in full calculations for \nucl{4}{\Lambda}{H} and \nucl{4}{\Lambda}{He}
and the CSB of the separation energies is also obtained from these full results. 
As can be seen in  Table~\ref{tab:csbpert0} for the $0^+$ state, the perturbative and 
full results agree well. We also checked that this is the case for the $1^+$ state. 
Additionally, the table shows the perturbatively estimated contributions of the kinetic energy, 
the YN interaction and the change in CSB of the nuclear interaction compared to the nuclear core
(mostly due to the change of the Coulomb interaction and the contribution of the proton-neutron mass
difference due to the compression of the core in the hypernucleus \cite{Bodmer:1985km}). 
It can also be seen that the major part of the CSB is due to the YN interaction. 
For the $0^+$ state, also the kinetic energy contribution due to the $\Sigma^+$ and $\Sigma^-$ 
mass difference is visible. This part is strongly related to the probability  to find a $\Sigma$
in \nucl{4}{\Lambda}{He} \cite{Nogga:2001ef}. It is therefore not surprising that its size 
is different for NLO13 and NLO19. This also shows explicitly that the individual contributions to CSB are not observable but depend on the realization of the YN interaction. 

\begin{table}[tbp]
\label{tab:scattlength}
\centering
\begin{tabular}{c|cc|cc|ccc}
&
$a^{\Lambda p}_{s}$ & $a^{\Lambda p}_{t}$ &
$a^{\Lambda n}_{s}$ & $a^{\Lambda n}_{t}$ &
$\chi^2(\Lambda p)$ & $\chi^2(\Sigma N)$ & $\chi^2({\rm total})$ \\
\hline
 NLO13(500) &-2.604 &-1.647 & -3.267 &-1.561 & 4.47 &12.13 & 16.60  \\
 NLO13(550) &-2.586 &-1.551 & -3.291 &-1.469 & 3.46 &12.03 & 15.49  \\
 NLO13(600) &-2.588 &-1.573 & -3.291 &-1.487 & 3.43 &12.38 & 15.81  \\
 NLO13(650) &-2.592 &-1.538 & -3.271 &-1.452 & 3.70 &12.57 & 16.27  \\
\hline
 NLO19(500) &-2.649 &-1.580 & -3.202 &-1.467 & 3.51 &14.69 & 18.20 \\
 NLO19(550) &-2.640 &-1.524 & -3.205 &-1.407 & 3.23 &14.19 & 17.42 \\
 NLO19(600) &-2.632 &-1.473 & -3.227 &-1.362 & 3.45 &12.68 & 16.13 \\
 NLO19(650) &-2.620 &-1.464 & -3.225 &-1.365 & 3.28 &12.76 & 16.04  \\
\end{tabular}
\caption{Singlet ($a^{\Lambda p/\Lambda n}_{s}$) and triplet ($a^{\Lambda p/\Lambda n}_{t}$) 
$S$-wave scattering lengths for $\Lambda p$ and $\Lambda p$ scattering in fm$^{-1}$. 
$\chi^2$ values for the comparison to the available YN data are also shown. 
}
\end{table}

Finally, we used the so determined CSB YN interactions to predict 
the scattering lengths for $\Lambda$N scattering. For the CSB scenario used 
here, it turns out that the CSB is small for the triplet and sizable for the singlet state. 
Interestingly, the dependence on the realization of the chiral interaction, NLO13 or NLO19,
and on the cutoff are small. Based on the central values of the currently best experimental 
results for the CSB of $A=4$ nuclei, we find that the singlet scattering lengths 
for $\Lambda p$ and $\Lambda n$ are $-2.6$~fm and $-3.2$~fm. This does not yet include 
the uncertainty due to the uncertainty of the experimental results for $A=4$ hypernuclei. 
Interestingly, the $\chi^2$ values with respect to the YN scattering data slightly decrease 
when CSB is used for the calculation compared to the isospin conserving case (see \cite{Haidenbauer:2019boi}). 
The stability of these predictions for the scattering lengths indicate
that the missing higher order interactions do not affect this result very much. 
In view of upcoming improved data for $A=4$ hypernuclei, an accurate determination 
of the $\Lambda$n scattering length will therefore be possible. 

\section{Conclusions and outlook}
\label{sec:concl}

In conclusion, we have shown that $A=4$ hypernuclei can provide important additional 
constraints for the $\Lambda p$ and $\Lambda n$ systems and can be used to accurately 
determine the CSB of these interactions. For the first time based on data, we found 
that the $\Lambda n$ interaction becomes more attractive in the \pw{1}{S}{0} partial wave 
when CSB is taken into account. For the currently best experimental data for $A=4$
hypernuclei, the magnitude of the singlet scattering length increases 
from $-2.9$~fm to $-3.2$~fm. In this scenario, the triplet scattering lengths 
is less affected by CSB. We note that we have studied other scenarios using older 
data in Ref.~\cite{Haidenbauer:2021wld}. For these scenarios, the changes of the scattering 
lengths can be different. It is therefore of utmost importance that also the 
\nucl{4}{\Lambda}{He} ground state energy and the \nucl{4}{\Lambda}{H} excitation energy 
is remeasured in order to confirm the old data and in order to reduce the experimental 
uncertainty. The work reported here shows that such data will provide clear constraints 
on the CSB of the YN interaction. 

But our calculations  should also be further refined.  First of all, chiral 3BFs \cite{Petschauer:2015elq}
should be included and could be used to improve the description of $A=4$ separation 
energies. With such improved interactions, the CSB of $p$-shell hypernuclei can 
be studied using the Jacobi no-core shell model \cite{Liebig:2015kwa,Le:2019gjp}.
This could open a path towards other, independent confirmations of 
the CSB in YN interactions. Further, the uncertainty of the experimental 
input should be taken into account to get a realistic estimate of the accuracy of 
our predictions. 

\acknowledgments

This work is supported in part by the DFG and the NSFC through
funds provided to the Sino-German CRC 110 ``Symmetries and
the Emergence of Structure in QCD'' (DFG grant. no. TRR~110)
and the VolkswagenStiftung (grant no. 93562) and 
has received funding from the European Research Council (ERC) under the European Union's Horizon 2020 research and innovation programme (grant agreement No. 101018170).
The work of UGM was supported in part by The Chinese Academy
of Sciences (CAS) President's International Fellowship Initiative (PIFI)
(grant no.~2018DM0034). We also acknowledge support of the THEIA net-working activity 
of the Strong 2020 Project. The numerical calculations were performed on JURECA
and the JURECA-Booster of the J\"ulich Supercomputing Centre, J\"ulich, Germany.

\bibliographystyle{JHEP}
\bibliography{bib/bbint,bib/tbfs,bib/ncsm,bib/hypexp,bib/nucmat,bib/srg,bib/fy,bib/methods,bib/astro}

\end{document}